\documentclass[prl,showpacs,twocolumn]{revtex4}
\usepackage{amsmath}
\usepackage{amssymb}
\usepackage{amscd}
\usepackage{amsthm}
\usepackage[ansinew]{inputenc}
\usepackage[T1]{fontenc}
\usepackage{ae,aecompl}
\usepackage[dvips]{graphicx}
\usepackage{pslatex}
\DeclareGraphicsExtensions{.eps}
\newcommand{\rmi}{{ i }}
\newcommand{\rme}{{ \rm e }}
\newcommand{\rmd}{{ \rm d }}
\newcommand{\nn}{\nonumber}
\newcommand{\n}{n}
\newcommand{\J}{v}
\newcommand{\U}{c}
\newcommand{\g}{c}
\newcommand{\s}{s}

\begin{document}
\bibliographystyle{apsrev}
\title[]{Mean-field dynamics of a non-hermitian Bose-Hubbard dimer}
\author{E. M. Graefe${}^1$}
\author{H. J. Korsch${}^1$}
\author{A. E. Niederle${}^{1,2}$}
\address{${}^1$ FB Physik, TU Kaiserslautern, D--67653 Kaiserslautern, Germany\\
${}^2$ Theoretical Physics, Saarland University, D--66041 Saarbr�cken, Germany}
\date{\today }

\begin{abstract}
We investigate an $N$-particle Bose-Hubbard dimer with an additional
effective decay term in one of the sites. A mean-field approximation
for this non-hermitian many-particle system is derived, based on a
coherent state approximation. The resulting nonlinear, non-hermitian two-level dynamics, 
in particular the fixed point structures showing characteristic modifications of the
self trapping transition, are analyzed. The mean-field dynamics is found to be in
reasonable agreement with the full many-particle evolution.
\end{abstract}


\maketitle

In the theoretical investigation of Bose-Einstein condensates (BEC)
the mean-field approximation leading to the description
via a Gross-Pitaevskii nonlinear Schr\"odinger equation (GPE) is
almost indispensable. It is usually achieved by replacing the
bosonic field operators in the multi-particle system with c-numbers
(the effective single-particle condensate wave functions), and
describes the system quite well for large particle numbers and low
temperatures. This approach is closely related to a classicalization
\cite{VardAnglBene} and allows for the application of
semiclassical methods \cite{semi}.

Recently considerable attention has been paid to effective
non-hermitian mean-field theories describing the scattering and
transport behavior of BECs
\cite{nlres}, as well as the
implications of decay (boundary dissipation)
\cite{06nlnh,LiviFran,Ng08}. The latter is closely related to atom
laser, for which it is possible to go beyond the mean-field
approximation and calculate the eigenmodes using Fano
diagonalization \cite{AL}. For linear quantum systems, an effective
non-hermitian Hamiltonian formalism proved useful and instructive
for the description of open quantum systems in various fields of
physics. Non-hermitian Hamiltonians typically yield complex
eigenvalues whose imaginary parts describe the rates with which an
eigenstate decays to the external world. Other kinds of
non-hermitian (PT-symmetric) quantum theories have also been
suggested as a generalization of quantum mechanics on a
fundamental level \cite{Bend07}.

However, the non-hermitian GPE has been formulated in an \textit{ad
hoc} manner as a generalization of the mean-field Hamiltonian and a
derivation starting from a non-hermitian many particle system is
required. This is as well interesting in a wider context of the classical limits of
effective non-hermitian quantum theories. In the present letter we
therefore introduce a generalized mean-field approximation and
investigate the characteristic features of the dynamics resulting
from the interplay of nonlinearity and non-hermiticity for a simple
many-particle Hamiltonian of Bose-Hubbard type, describing a BEC in
a leaking double well trap:
\begin{eqnarray}
\hat{\cal H}&=&(\epsilon-2\rmi\gamma)\hat a_1^\dagger\hat a_1  -
\epsilon \hat a_2^\dagger\hat a_2
 + \J (\hat a_1^\dagger \hat a_2 + \hat a_1 \hat a_2^\dagger)
 \nonumber\\
 &&+\frac{\U }{2}(\hat a_1^\dagger\hat a_1  - \hat a_2^\dagger\hat
 a_2)^2.
 \label{BH-hamiltonian}
\end{eqnarray}
Here $\hat a_j$, $\hat a_j^\dagger$ are bosonic particle
annihilation and creation operators for the  $j$th mode. The onsite
energies are $\pm\epsilon$, $\J $ is the coupling constant and $\U $
is the strength of the onsite interaction. The additional imaginary
part of the mode energy $\gamma$ describes the
first mode as a resonance state with a finite lifetime, like, e.g.,
the Wannier-Stark states for a tilted optical lattice
\cite{02wsrep}. A direct experimental realization could be achieved
by tunneling escape of atoms from one of the wells. Even in the
non-hermitian case, the Hamiltonian commutes with the total number
operator $\hat N=\hat a_1^\dagger\hat a_1  + \hat a_2^\dagger\hat
a_2$ and the number $N$ of particles is conserved. The ``decay''
describes not a loss of particles but models the decay of the
probability to find the particles in the two sites considered here.

First theoretical results for the spectrum of the non-hermitian
two-site Bose-Hubbard system (\ref{BH-hamiltonian}) and a closely
related PT-symmetric system were presented in \cite{Hill06,08PT}. In
this paper we will present first results for the {\it dynamics\/} of
this decaying many-particle system with emphasis on the mean-field
limit of large particle numbers. In order to specify the mean-field
approximation in a controllable manner, we derive coupled equations
for expectation values under the assumption that the system,
initially in a coherent state, remains coherent for all times of
interest. This is a direct extension of the frozen Gaussian
approximation in flat phase space \cite{Hell81, Kluk86} to $SU(2)$
coherent states, relevant to the present case as discussed below.
This yields classical evolution equations for the coherent states
parameters.

It facilitates the analysis to rewrite the Hamiltonian
(\ref{BH-hamiltonian}) in terms of angular momentum operators
$\hat L_x ={\textstyle \frac12}(\hat a_1^\dagger\hat a_2+\hat a_1\hat a_2^\dagger)$, $\hat L_y={\textstyle \frac{1}{2\rmi}}(\hat a_1^\dagger\hat a_2-\hat a_1\hat a_2^\dagger)$ and $\hat L_z={\textstyle \frac12}(\hat a_1^\dagger\hat a_1-\hat a_2^\dagger\hat a_2)$, 
satisfying the commutation rules $[\hat L_x,\hat L_z]=\rmi \hat L_z$
and cyclic permutations, as
\begin{eqnarray}
\hat{\cal H}=2 (\epsilon-\rmi \gamma )\hat L_z+2\J \hat L_x+2\U \hat
L^2_z-\rmi \gamma \hat N \,. \label{BH-hamiltonian-L}
\end{eqnarray}
The conservation of $\hat N$ appears as the conservation of $\hat
L^2=\frac{\hat N}{2}\big(\frac{\hat N}{2}+1\big)$, i.e. the
rotational quantum number $\ell =N/2$. The system dynamics is
therefore restricted to an $(N+1)$-dimensional subspace and can be
described in terms of the Fock states $|k,N-k\rangle$,
$k=0,\ldots,N$ or the $SU(2)$ coherent states \cite{Zhan90},
describing a pure BEC:
\begin{eqnarray}
|x_1,x_2\rangle &=& \frac{1}{\sqrt{N!}}\left(x_1 \hat
a_1^\dagger+x_2 \hat a_2^\dagger \right)^N |0\rangle,
\label{cohstate}
\end{eqnarray}
with $x_j\in \mathbb C$. The norm, which may differ from unity, is
$\langle x_1,x_2|x_1,x_2\rangle =\n^N$, where $\n = | x_1 |^2 + | x_2 |^2$.

A general discussion of the time evolution of a quantum system under
a non-hermitian Hamiltonian $\hat{\cal H}=\hat H -\rmi \hat\Gamma$
with hermitian $\hat H$ and $\hat\Gamma$ can be found in
\cite{Datt90b}. Matrix elements of an operator $\hat A$ without
explicit time-dependence satisfy the generalized Heisenberg
equation, which in our case becomes
\begin{eqnarray}
\nn \rmi \hbar \frac{\rmd \,}{\rmd\, t}\langle \psi|\hat A|\psi
\rangle &=&\langle \psi|\hat A\hat{\cal H}-\hat{\cal H}^\dagger\hat
A|\psi \rangle\\ &=&\langle \psi|\,[\hat A,\hat H]|\psi \rangle -
i\langle \psi|\,[\hat A,\hat \Gamma]_{\scriptscriptstyle +}|\psi
\rangle, \label{dattoli1}
\end{eqnarray}
where $[\cdots ]_{\scriptscriptstyle +}$ is the anti-commutator. As an
immediate consequence of the non-hermiticity, the norm of the
quantum state is not conserved, $\hbar \frac{\rmd \,}{\rmd\,
t}\langle \psi|\psi \rangle =-2\langle \psi|\hat \Gamma|\psi \rangle
\,$, thus the survival probability decays exponentially for the
simple case of a constant $\Gamma>0$. The time evolution of the
expectation value of an observable $\langle \hat A\rangle =\langle
\psi|\hat A|\psi \rangle/\langle \psi|\psi \rangle$ is described by
the equation of motion
\begin{eqnarray}\label{HeisenbergEnde}
\rmi \hbar \frac{\rmd \,}{\rmd\, t}\langle \hat A \rangle
=\langle[\hat A,\hat H]\rangle - 2\rmi \,\Delta^2_{A\Gamma} \,,
\label{expect-t}
\end{eqnarray}
with the covariance $\Delta^2_{A\Gamma}=\langle {\textstyle
\frac12}[ \hat A, \hat \Gamma]_{\scriptscriptstyle +} \rangle -
\langle  \hat A \rangle \langle \hat \Gamma \rangle\,.$

For the Bose-Hubbard system (\ref{BH-hamiltonian-L}) these evolution
equations, formulated in terms of the angular momentum operators
with read (units with $\hbar=1$
are used in the following)
\begin{eqnarray}\label{KomZerfall1}
{\textstyle \frac{{\rmd}}{\rmd\, t}} \langle \hat L_x \rangle \!\!\!&=&\!\!\!
 - 2 \epsilon \langle \hat L_y \rangle - 2\U  \langle [ \hat L_y, \hat L_z]_{\scriptscriptstyle +} \rangle
 - 2\gamma  \,\lbrace 2 \Delta^2_{L_xL_z} \!+\! \Delta^2_{L_x,N}\rbrace  \nonumber\\[2mm]
{\textstyle \frac{\rmd}{\rmd\, t}} \langle \hat L_y \rangle \!\!\!&=&\!\!\!
  2\epsilon \langle \hat L_x \rangle + 2\U  \langle [ \hat L_x, \hat L_z]_{\scriptscriptstyle +} \rangle
-2\J  \langle \hat L_z \rangle - 2\gamma  \,\lbrace 2 \Delta^2_{L_yL_z} \!+\! \Delta^2_{L_y,N}\rbrace \nonumber\\[2mm]
{\textstyle \frac{\rmd}{\rmd\, t}} \langle \hat L_z \rangle \!\!\!&=&\!\!\!
  2 \J  \langle \hat L_y \rangle - 2 \gamma  \,\lbrace 2\Delta^2_{L_zL_z} \!+\! \Delta^2_{L_zN}\rbrace
\end{eqnarray}
and the norm decays according to
\begin{equation}
 \frac{\rmd}{\rmd\, t} \langle \psi| \psi \, \rangle = -2\gamma \,\big\lbrace 2\langle \hat L_z \rangle+ \langle \hat N \rangle \big\rbrace
 \langle \psi | \psi \, \rangle.
\end{equation}

In order to establish a mean-field description, we choose a coherent
initial state $|x_1,x_2\rangle$, i.e.~a most classical state, and
assume that it remains coherent for all times of interest. This
assumption is, in fact, exact, if the Hamiltonian is a linear
superposition of the generators of the dynamical symmetry group,
i.e.~for vanishing interaction $\U=0$ (the proof in \cite{Zhan90}
can be directly extended to the non-hermitian case). For the
interacting case $\U \neq0$ this is an approximation and the
mean-field equations of motion are obtained by replacing the
expectation values in the generalized Heisenberg equations of motion
(\ref{KomZerfall1}) with their values in $SU(2)$ coherent states
(\ref{cohstate}).

The $SU(2)$ expectation values of the $\hat L_i$, $i=x, y, z$, read
\begin{gather}
\label{Erwdreh}
\s_x =\frac{x_1^* x_2+ x_1 x_2^*}{2\n}\ ,\
\s_y=\frac{x_1^* x_2- x_1 x_2^*}{2\rmi n}\ ,\
\s_z =\frac{x_1^* x_1 -  x_2^* x_2}{2\n},
\end{gather}
with the abbreviations $\s_j=\langle \hat L_j \rangle/N$ for the
mean values per particle; the expectation values of the
anti-commutators factorize as
\begin{eqnarray} \label{ErwAntiDrehexakt}
\langle [\hat L_i, \hat L_j]_{\scriptscriptstyle +} \rangle = 2 \left(1-\frac{1}{N} \right) \langle \hat L_i \rangle \langle
\hat L_j \rangle + \delta_{ij} \frac{N}{2}\,,
\end{eqnarray}
and $\langle [\hat L_i, \hat N]_{\scriptscriptstyle +} \rangle = 2N\langle \hat L_i \rangle$.
Inserting these expressions into (\ref{KomZerfall1}) and taking the macroscopic limit
$N\to\infty$ with $N\U =\g $  fixed, we obtain the desired non-hermitian
mean-field evolution equations:
\begin{eqnarray}\label{KomZerfall2}
\begin{array}{rrrrll}
\dot{\s_x}=&-2 \epsilon \s_y&- 4 \g \s_z \s_y& &+ 4 \gamma \,\s_z \s_x, \\
\dot{\s_y}=&+2 \epsilon \s_x&+4 \g \s_z \s_x &-2 \J  \s_z &+ 4 \gamma \,\s_z \s_y,\\
\dot{\s_z}=& & &+2 \J  \s_y&-\gamma\, (1-4\s_z^2)\,.
\end{array}
\end{eqnarray}
These nonlinear Bloch equations are real valued and conserve
$\s^2=\s^2_x+\s^2_y+\s^2_z=1/4$, i.e.~the dynamics is regular and
the total probability $n$ decays as
\begin{equation}\label{Norm}
\dot n =- 2 \gamma \left(2\s_z+1\right) n\,.
\end{equation}

Equivalently, the nonlinear Bloch equations (\ref{KomZerfall2}) can
be written in terms of a non-hermitian generalization of the
discrete nonlinear Schr\"odinger equation, i.e. for the
time-evolution of the coherent state parameters $x_1$, $x_2$. Most
interestingly, these equations are canonical, $\rmi \dot
x_j=\partial H/\partial x_j^*$, $\rmi \dot x_j^*=-\partial
H^*/\partial x_j$, $j=1,2$, where the Hamiltonian function is
related to the expectation value of the Hamiltonian $\hat {\cal H}$:
$H(x_1,x_1^*,x_2,x_2^*)=\langle \hat{\cal H} \rangle n/N$ and can be
conveniently rewritten in terms of the quantities $\psi_j=\rme^{\rmi
\beta}x_j$ where the (irrelevant) total phase is adjusted according
to $\dot \beta =-g\kappa^2$ with $\kappa=(|\psi_1|^2-|\psi_2|^2)/n$.
The resulting discrete non-hermitian GPE reads
\begin{equation}\label{nlnhGP}
\rmi\frac{\rmd}{\rmd\, t}\begin{pmatrix} {\psi}_1 \\{\psi}_2 \end{pmatrix}
=\begin{pmatrix} \varepsilon + g \kappa - 2\rmi \gamma & v \\
   v & - \varepsilon - g\kappa  \end{pmatrix}
   \begin{pmatrix}\psi_1\\\psi_2\end{pmatrix}.
\end{equation}
Similar non-hermitian mean-field equations, with the
choice $\kappa=|\psi_1|^2-|\psi_2|^2$, leading to different
dynamics, have been suggested and studied before
\cite{Scha97b,Hill06,06nlnh,LiviFran}. These \textit{ad hoc}
nonlinear non-hermitian equations also appear for absorbing
nonlinear waveguides \cite{Muss08}.

\begin{figure}[b]
\centering
\includegraphics[width=4cm]{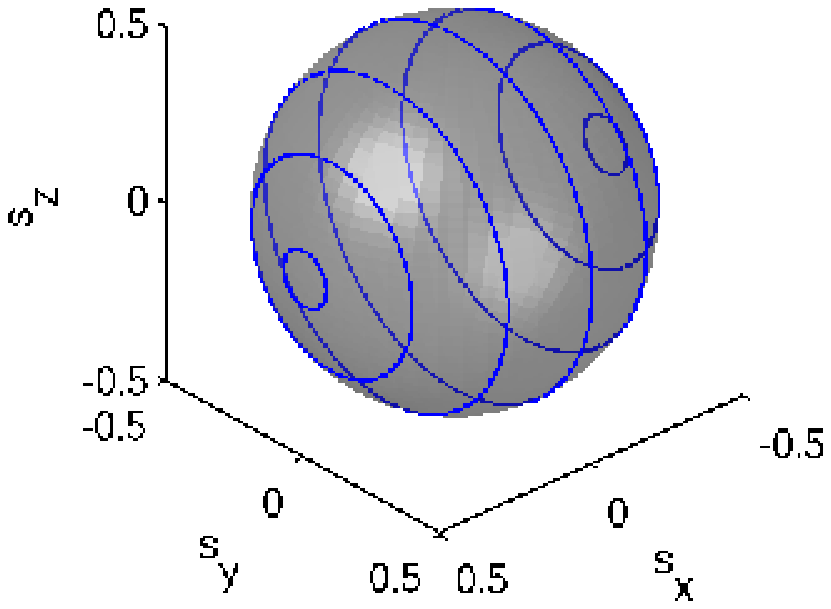}
\includegraphics[width=4cm]{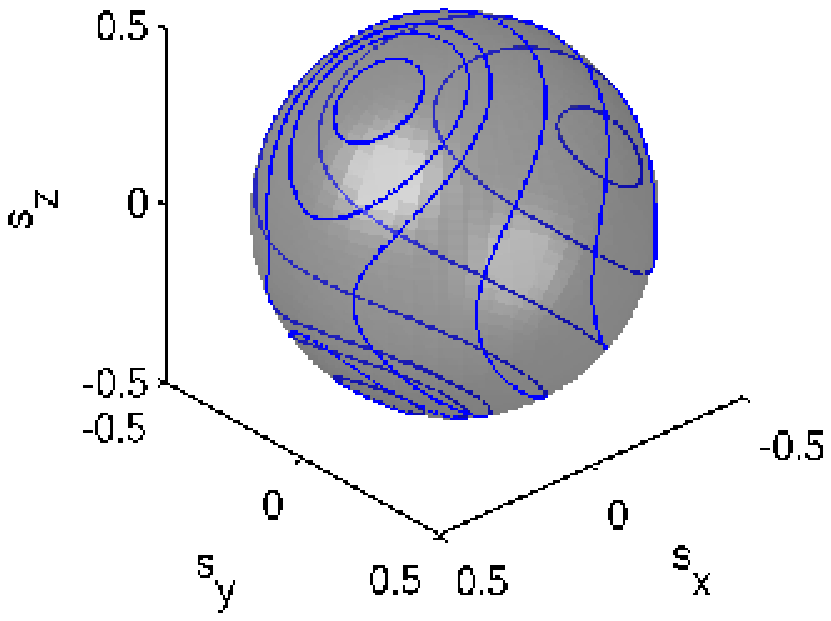}
\includegraphics[width=4cm]{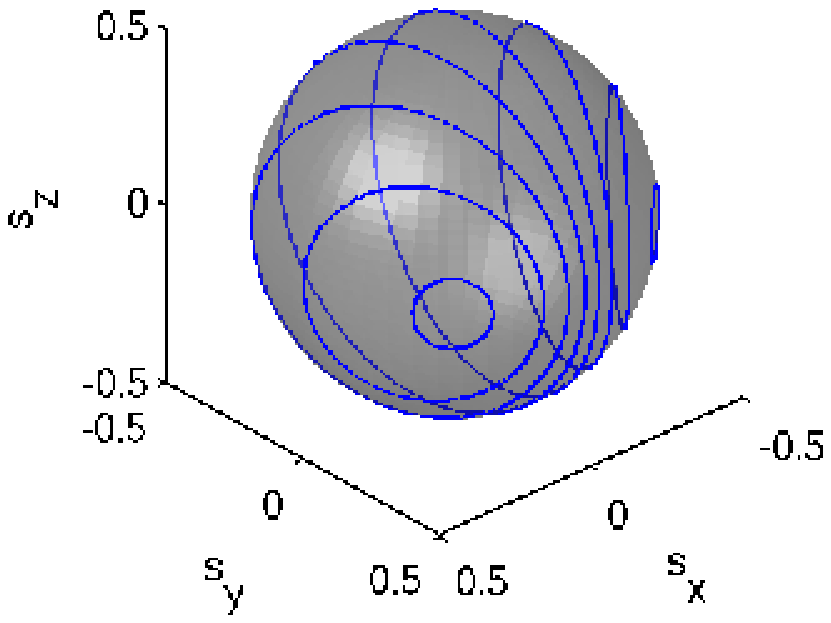}
\includegraphics[width=4cm]{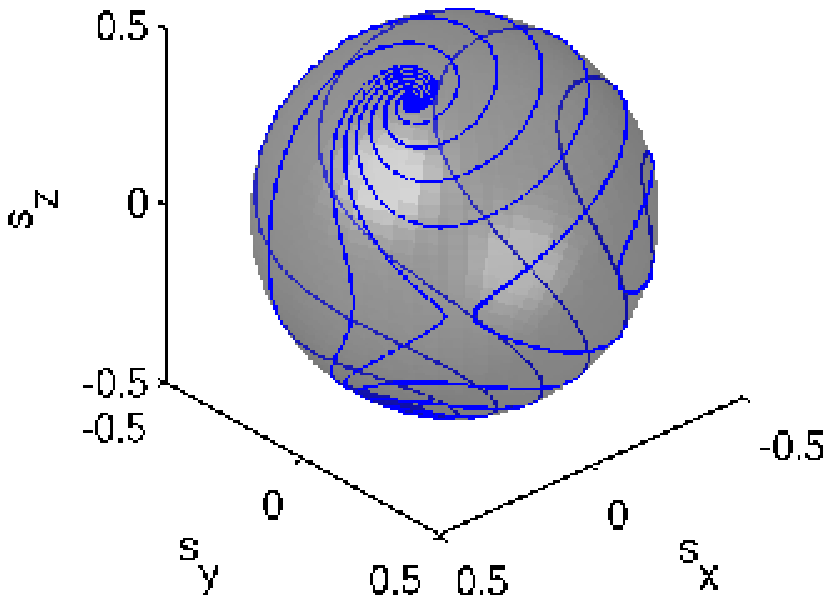}
\caption{\label{fig-phasespace}
(Color online) Mean-field dynamics on the Bloch sphere for the hermitian  $\gamma=0$ (top)
and the non-hermitian case $\gamma=0.75$ (bottom) for $\g=0$ (left) and $\g=2$ (right) and $\epsilon=0$ and $\J=1$.}
\end{figure}

The dynamics of the  nonlinear Bloch equations (\ref{KomZerfall2})
is organized by the fixed points which are given by the real roots
of the fourth order polynomial:
\begin{equation}\label{fixedpoints}
4(\g^2+\gamma^2)\s_z^4+4\g\epsilon \s_z^3+(\epsilon^2+\J^2 -\g^2-\gamma^2)\s_z^2-\g\epsilon\s_z-\epsilon^2/4=0\,.
\end{equation}
In the following we will restrict ourselves to the symmetric case
$\epsilon=0$. Then the polynomial (\ref{fixedpoints}) becomes
biquadratic and the fixed points are easily found analytically.

In parameter space we have to distinguish three different regions:
(a) For $\g^2+\gamma^2<\J^2$, we have two fixed points which are
both simple centers. (b) For $|\gamma| > |\J|$, we have again two
fixed points, a sink and a source. (c) Four coexisting fixed points
are found in the remaining region, namely a sink and a source (respectively
two centers for $\gamma=0$), a center and a saddle point. Note that
the index sum of these singular points on the Bloch sphere must be
conserved under bifurcations and equal to two \cite{Arno06}.
Bifurcations occur at critical parameter values: For
$\g^2+\gamma^2=\J^2$ (and $\gamma\neq0$), one of the two centers
(index $+1$) bifurcates into a saddle (index $-1$) and two foci
(index $+1$), one stable (a sink) and one unstable (a source). This
is a non-hermitian generalization of the selftrapping transition for
$\gamma=0$. The corresponding critical interaction strength is
decreased by the non-hermiticity, i.e. the decay supports
selftrapping . For $\gamma=\pm \J$, the saddle (index $-1$) and the
center (index $+1$) meet and disappear. For $g=0$, we observe a
non-generic bifurcation at $\gamma=\pm\J$ (an exceptional point
\cite{08PT}) where the two centers meet and simultaneously change
into a sink and a source.

\begin{figure}[t]
\centering
\includegraphics[width=4cm]{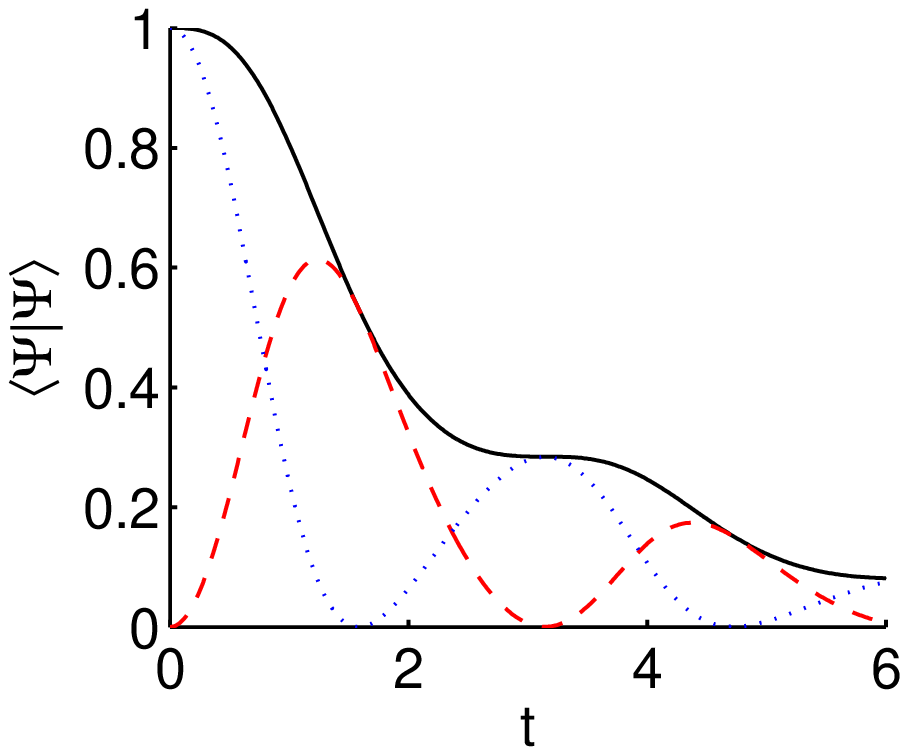}
\includegraphics[width=4cm]{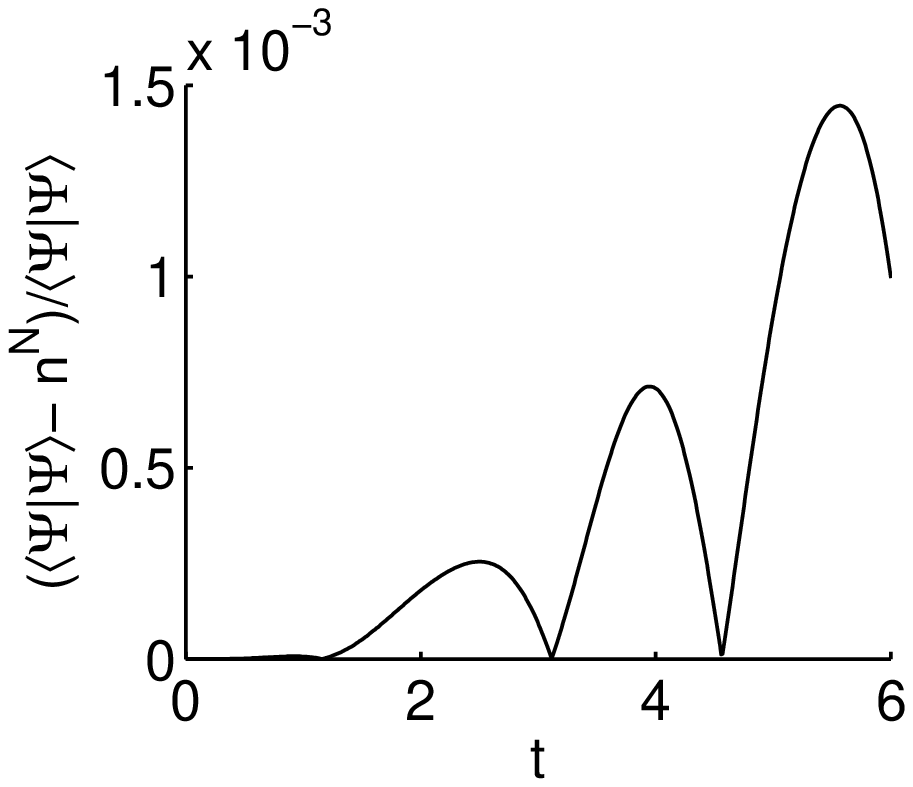}
\caption{\label{fig-comp1}
(Color online) Decay of the survival probability (full black curve) and the populations of site 1 (dashed red curve) and 2 (dotted blue curve) for an initial coherent state located at the south pole, for $\g=0.1$, $\gamma=0.01$, $\J=1$ and $N=20$ (left) and the relative deviations between many-particle and mean-field results (right).}
\end{figure}

As an example, Fig.~\ref{fig-phasespace} shows the flow
(\ref{Erwdreh}) on the Bloch sphere for $\J=1$ both for the
hermitian $\gamma=0$ (top) and the non-hermitian case $\gamma=0.75$
(bottom).  For $\gamma=0$ we observe the well-known selftrapping
effect: In the interaction free case $\g=0$ (upper left) we have two
centers at $\s_y=\s_z=0$, $\s_x=\pm \frac{1}{2}$ and Rabi
oscillations. Increasing the interaction $\g$ one of the centers
bifurcates into a saddle (still at $\s_z=0$) and two centers, which
approach the poles (upper right for $\g=2$). The corresponding
nonlinear stationary states therefore favor one of the wells. In the
decaying system with $\gamma=0.75$ (bottom), these patterns are
changed. For  $\g=0$ (lower left) we are still in region (a) with
two centers located on the equator; however, they move towards
$s_x=0,\ s_y=\frac{1}{2}$, approaching each other. For $\g=2$ (lower
right), in region (b) above the bifurcation, we have a center, a
sink (lower hemisphere), a source (upper hemisphere) and a saddle.
The system relaxes to a state with excess population in the
non-decaying well, i.e. the selftrapping oscillations are damped,
which is in agreement with the effect of decoherence in a related
nonlinear two mode system reported in \cite{Wang07}. Finally, in
region (c) only a source and a sink survive and the flow pattern
simplifies again (not shown). The manifestation of the different
mean-field regimes in the many particle system is the occurrence and
unfolding of higher order exceptional points in the spectrum
\cite{08PT}.

Let us finally compare the mean-field evolution with the full
many-particle dynamics. The full quantum solution is obtained by
numerically integrating the Schr\"odinger equation for the
Bose-Hubbard Hamiltonian (\ref{BH-hamiltonian}) for an initial
coherent state with unit norm. Figure \ref{fig-comp1} shows the
decay of the total survival probability $\langle\psi|\psi\rangle$ as
a function of time for weak interaction ($\g=0.1$) and weak decay
($\gamma=0.01$) with $\J=1$, when initially the non-decaying site
$2$ is populated. The multi-particle results agree with the
mean-field counterpart $n^N$ on the scale of drawing. The deviation
increases with time as can be seen on the right side. The
probability shows a characteristic staircase behavior (see also
\cite{LiviFran, Ng08}) due to the fact that the population
oscillates between the two sites and the decay is fast when site 1
is strongly populated and slow if it is empty. This picture is
confirmed by the populations $\langle \psi|\hat a_1^\dagger\hat
a_1|\psi\rangle/N$ and $\langle\psi| \hat a_2^\dagger\hat
a_2|\psi\rangle/N$ of the two sites also shown in the figure. These
quantities agree with their mean-field counterparts
$(1/2+\s_z)n^N/2$ and $(1/2-\s_z)n^N/2$ on the scale of drawing.
The overall decay of the norm is approximately exponential,
$\frac{\rmd}{\rmd\, t} \langle \psi|\psi\rangle\approx -2 \gamma
N\langle \psi|\psi\rangle$ within region (a), as seen from
(\ref{Norm}) with $\overline{\s_z}=0$.

\begin{figure}[t]
\centering
\includegraphics[width=4cm]{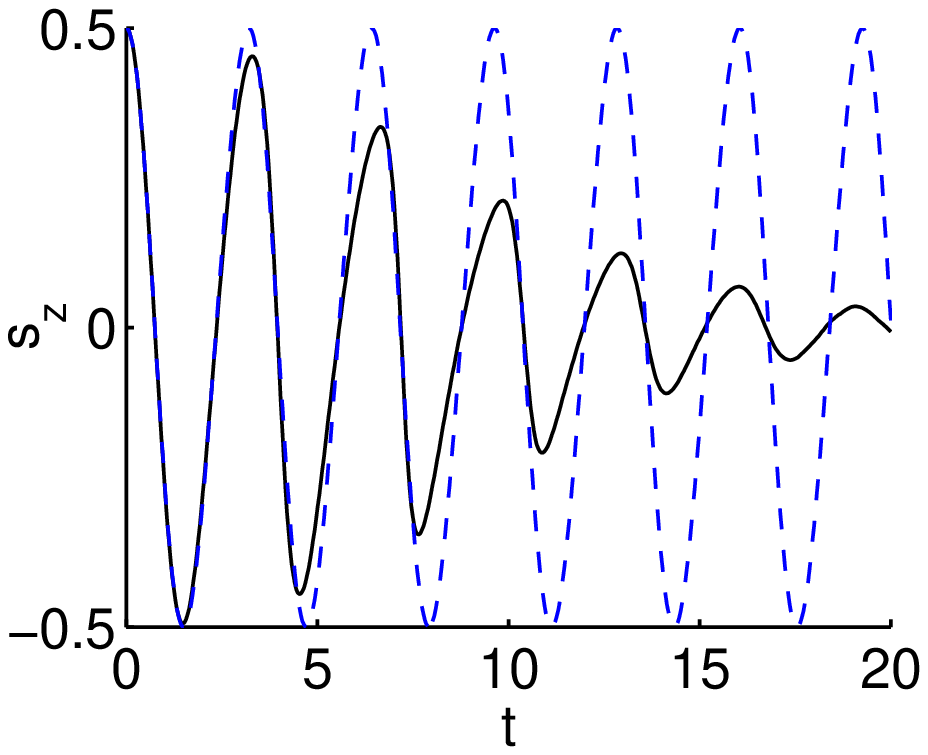}
\includegraphics[width=4cm]{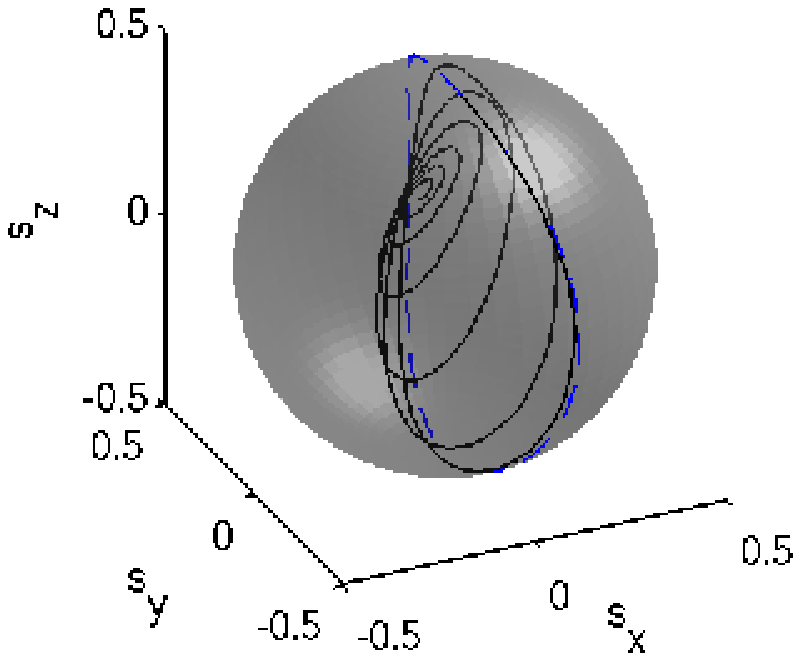}
\includegraphics[width=4cm]{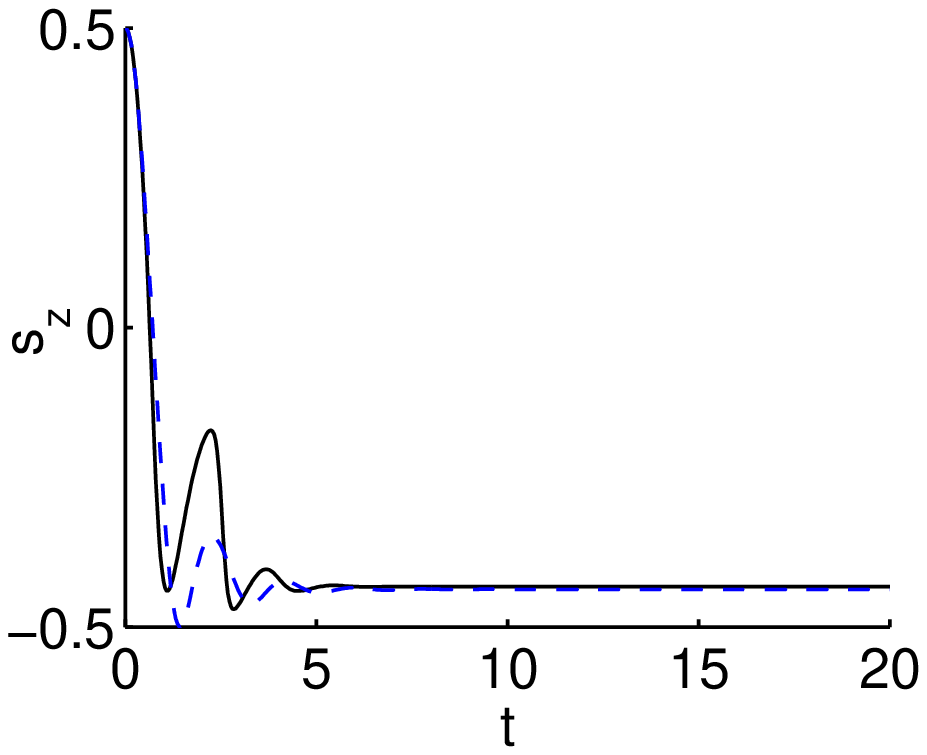}
\includegraphics[width=4cm]{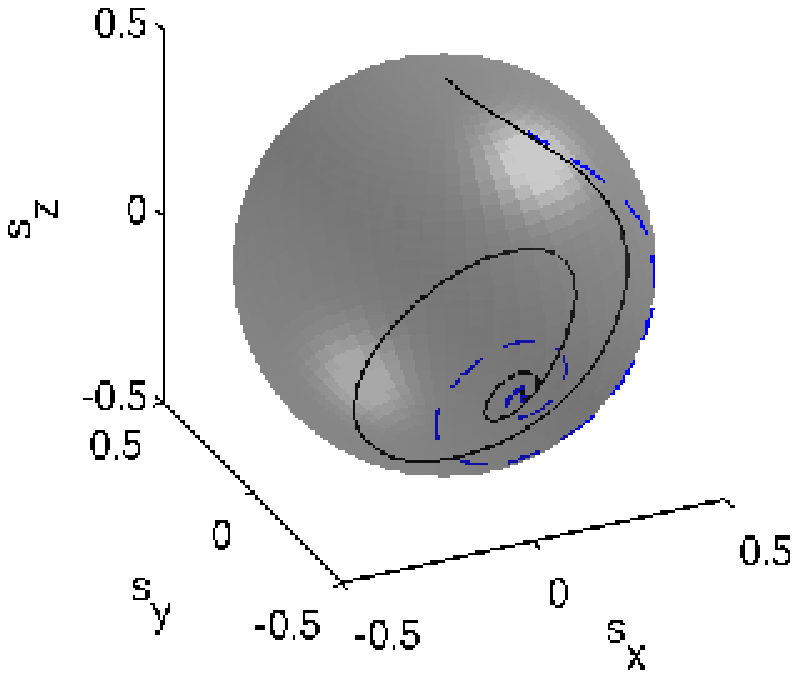}
\caption{\label{fig-comp2}
(Color online) Mean-field evolution of the population imbalance $\s_z(t)$ (dashed blue curve)
in comparison with the full many particle system for $N=20$ particles (black curve)
and an initial coherent state located at the north pole ($\U=0.5$, $\gamma=0.1$ (top) and $\U=2$, $\gamma=0.5$ (bottom) and $\J=1$). }
\end{figure}

The dynamics on the Bloch sphere in region (a) typically show
Rabi-type oscillations. An example with parameters  $\g=0.5$ and
$\gamma=0.1$ is shown in Fig.~\ref{fig-comp2}. The 
mean-field oscillation follows a big loop extending over the whole
Bloch sphere. The many-particle motion oscillates with the same
period, however, with a decreasing amplitude. This effect, known as
breakdown of the mean-field approximation in the hermitian case, is
due to the spreading of the quantum phase space density over the
Bloch sphere, and can be partially cured by averaging over a density
distribution of mean-field trajectories \cite{07phase12}.

For strong interaction, i.e.~in the selftrapping region (c), we find
an attractive fixed point, a sink, in the mean-field dynamics. An
example is shown in Fig.~\ref{fig-comp2} for $\g=2$ and
$\gamma=0.5$. The mean-field trajectory, which started at the north pole,
approaches the fixed point at $\s_{z,0}=-0.433$. The full
many-particle system shows a very similar behavior.

Further numerical investigations show that the short time behavior
of the many-particle dynamics, as well as characteristic quantities
such as, e.g., the half-life time, are extremely well captured by
the mean-field description in most parameter ranges.

In this letter, we have constructed a mean-field approximation for a
non-hermitian many-particle Hamiltonian, which can directly be generalized to
other effective non-Hermitian Hamiltonians. The resulting dynamics
differ from the \textit{ad hoc} non-hermitian evolution equations
used in previous studies. It should be noted that the nonlinear Bloch
equations (\ref{KomZerfall2}) can be derived in an alternative way,
based on a recently formulated number-conserving evolution equation
in quantum phase space for $M$-site Bose-Hubbard systems
\cite{07phase12}, which allows for an immediate extension to the
non-hermitian case.

Support from the DFG via the GRK 792 is gratefully acknowledged. We
thank Friederike Trimborn and Dirk Witthaut for fruitful and
stimulating discussions.

\end{document}